# The effect of toroidicity on relaxed states at low aspect ratio

R. Paccagnella[1] , S. Masamune[2] , I. Predebon[1] , A. Sanpei[2]

[1] Consorzio RFX and C.N.R , 35127 Padova, Italy

[2] Kyoto Institute of Technology, 606-8585 Kyoto, Japan

**Abstract**

The experimentally measured single helical states in the low aspect ratio (R/a=2) RELAX device have been compared with the predictions of a cylindrical relaxation theory and a substantial deviation in terms of some macroscopic characteristics has been found, in particular for cases corresponding to a large reversal of the edge toroidal field.

In this paper we consider the effect of the toroidal geometry by employing the VMEC and the RelaxFit equilibrium codes. We show that the inclusion of toroidicity produces significant corrections to the cylindrical states, especially for the deep reversed cases.

**Introduction:**

The Reversed Field Pinch (RFP) is a low safety factor magnetic confinement device where the equilibrium magnetic field lines describe helices that turn several times around the poloidal angle while encircling once the torus. For this reason the toroidal effects are generally small since, for example, particles moving along the field lines are not much affected by the toroidal curvature. However, beside RFP stability which is not much affected by toroidicity [1,2], important exceptions are the transport properties (not considered here) and the plasma equilibrium, as we will discuss in this paper.

The classical relaxation theory, in cylindrical geometry, proposed several years ago by J.B.Taylor [3] has been able to give a surprisingly simple explanation for the observed RFP equilibrium states, in particular the reversal of the toroidal magnetic field at the plasma edge.

In more recent years, several RFP research studies have been dedicated to the interpretation of the experimental observation of states where a single magneto-hydro-dynamic (MHD) mode is dominating the fluctuations spectrum. These plasmas have been named Single Helical (SH) or Quasi Single Helical (QSH), depending on their spectral characteristics and to distinguish them from the Multi-Helical (MH) cases, where many MHD modes are simultaneously excited [4-8]. In the past and present experiments, the SH (or QSH) states have been routinely detected, but generally observed to be oscillating in time, never reaching a steady state; however it is believed that they could represent a tendency of the plasma to relax towards a helical equilibrium.

In a recent paper [9] a relaxation theory [10] has been compared with the experimental data obtained in the low aspect ratio RELAX device [11]. The theory is applicable to SH cases observed also in RELAX itself (see also Refs. 5 to 10 in [9]).

It should be noted that, although the relaxation theory [10] assumes the presence of a dominant m/n mode, the form of the related eigenfunction remains completely undetermined and the resulting equilibria are cylindrical. Indeed, within the purpose of the present paper the helical correction is not important (see the Discussion), as its effect on averaged quantities (like F and Θ, see below) is negligible. In addition, the experimental toroidally-averaged measurements are also insensitive to helical corrections, while they are clearly affected by the toroidal geometry. Therefore some deviation from the cylindrical theoretical predictions of the measurements could be expected. The purpose of the present paper is to quantify this effect.

The cylindrical equilibria deduced from the relaxation theory, that we indicate in the following as Single Helical Relaxed (SHR) states, however deviate from the experimental F-Θ curves (see section 1 for definitions), especially for deep reversed toroidal field cases at high Θ values. In Ref.[9], it was speculated that this discrepancy could be explained either by physical effects, due to a residual stochasticity in the core leading to a flattening of the current profile, or by toroidal geometrical effects which are expected to be particularly strong at low aspect ratio. The aim of the present paper is to test more quantitatively the second hypothesis. Unfortunately the relaxation theory cannot be extended directly to toroidal geometry, because the problem becomes mathematically intractable, therefore we will employ a different strategy by using two toroidal codes, VMEC [12] and RelaxFit [13], and then compare their predictions with those obtained from the cylindrical SHR states.

In the context of this paper, the code VMEC will be used to solve the equilibrium problem in fixed boundary mode, by prescribing the safety factor q profile as a function - in the RFP version of VMEC - of the normalized poloidal flux; furthermore, the code will be used with imposed axisymmetry, i.e., without 3D displacements of the magnetic axis. With the aim to investigate the role of toroidicity on the equilibrium field, VMEC will be at first initialized with the safety factor obtained from the SHR states.

On the other hand, the RelaxFit code is a Grad-Shafranov equation solver which uses the magnetic measurements at the wall to reconstruct the plasma equilibrium as a free boundary problem. Note however that in this paper we assume a fixed-boundary at the shell radius. In this case, the initialization of VMEC with the RelaxFit output safety factor allows us to benchmark the two toroidal codes, which adopt quite different strategies and assumptions for the solution of the plasma equilibrium.

The paper is organized as follows: in section 1 a brief summary of the relaxation theory is given and some key parameters are discussed; in section 2 the RelaxFit code and the equilibrium reconstruction method is presented; in section 3 a few useful information about the VMEC code are described; in section 4 the toroidal equilibria obtained with VMEC are compared with some cylindrical states; in section 5 the RelaxFit output is used to initialize VMEC and a cross-comparison of the two toroidal codes is done; moreover the VMEC changes to the SHR states are analyzed more specifically. Finally, a discussion and conclusions are given.

## 1. Single Helical relaxed states

As discussed in detail in Refs. [9] and [10], the Taylor's relaxation theory [3] can be modified in the case of a dominant tearing mode.

In particular it was shown that the two invariants:

$$K_o = \frac{1}{2}\int_V \boldsymbol{A} \cdot \boldsymbol{B} \, dV$$

which is the Taylor's original total helicity invariant, and :

$$K_1 = \frac{1}{2}\int_V \chi \, \boldsymbol{A} \cdot \boldsymbol{B} \, dV$$

which is a "weighted" total helicity, are well conserved [10]. The weight in $K_1$ is the helical flux function: $\chi = q_S \Psi - \Phi$ where $q_S = m/n$ is the pitch of the dominant mode (m and n being respectively the poloidal and toroidal mode numbers), $\Psi$ and $\Phi$ are the poloidal and toroidal fluxes.

In a second paper [14] in order to preserve gauge invariance, a hierarchy of invariants was introduced:

$$K_d = \frac{1}{2}\int_V \chi^d \mathbf{A} \cdot \mathbf{B}\, dV$$

where d is an arbitrary null or positive integer . For d=0 the $K_0$ invariant is recovered.

Assuming the boundary condition $\lambda(a)=0$ (where $\lambda = \mathbf{J}\,\mathbf{B}/B^2$ is the current component parallel to the magnetic field ) the relaxed state is found to be:

$$\mathbf{J} = \sum_d \frac{\lambda_d\,(d+2)}{2} \chi^d \mathbf{B} \qquad (1)$$

where $\lambda_d$'s are eigenvalues.

At difference with the Taylor's theory that predicts a constant $\lambda$ profile, the interesting thing here is that the $\lambda$ profile varies with the radius according to Eq.(1). Note also that the mode helicity (pitch), which is a free parameter of the theory, is influencing the final states since it enters the definition of the function $\chi$. In this paper we will consider only the first two terms (d=0,1) in the sum of Eq.(1), which is equivalent to assume $K_o$ and $K_1$ as the only conserved quantities of the hierarchy. The SHR states described by Eq.(1) are cylindrical symmetric.

RFP relaxed states are usually characterized by two non-dimensional parameters [3]:

$$F = \frac{B_z(a)}{<B_z>} \quad \text{with} \quad <B_z> = \frac{2}{a^2}\int_0^a B_z\, r\, dr \qquad (2a)$$

and

$$\Theta = \frac{B_\theta(a)}{<B_z>} \qquad (2b)$$

i.e. the ratios between the toroidal and poloidal fields at the wall to the average toroidal field respectively. It was shown, for constant $\lambda$, that $\lambda a = 2\Theta$ (a being the plasma minor radius) [3].

The definitions (2a,2b) are valid in the cylindrical symmetry of Eq.(1). For the toroidal case they need to be generalized as, for example, in [1]. In this paper we will use definitions suitable for a direct comparison with the experimental measurements of F and $\Theta$ (see Section 3).

## 2. Experimental equilibrium reconstruction with the RelaxFit code

The RelaxFit [13] code is an axi-symmetric toroidal equilibrium reconstruction code, which was modified from the MSTFit [15] code to be applied to the low aspect ratio RELAX device [11]. The code consists in a solver of the Grad-Shafranov (GS) equation for the poloidal flux function ψ, and in a optimization procedure to reconstruct the equilibrium with maximum consistency with experimental observations.

The GS equation is a second-order nonlinear partial differential equation for the poloidal flux function ψ(R,Z) which describes an axisymmetric toroidal equilibrium,

$$\left[R\frac{\partial}{\partial R}\left(\frac{1}{R}\frac{\partial}{\partial R}\right) + \frac{\partial^2}{\partial Z^2}\right]\psi = \Delta^*\psi = -\mu_0 R J_\phi$$

$$J_\phi(R,Z) = \frac{2\pi F(\psi)F'(\psi)}{\mu_0 R} + 2\pi R p'(\psi) \qquad (3)$$

where $J_\phi$ is the toroidal current density, $F(\psi)=RB_\phi(R,Z)$ the poloidal current function with $B_\phi$ the toroidal magnetic field, and $p(\psi)$ the pressure. F(ψ) and p(ψ) are specified through a set of free parameters (see below) which define their functional forms. In the optimization procedure the set of parameters are varied to obtain the optimal equilibrium solution consistent with the magnetic measurements.

Once the functional forms of F(ψ) and p(ψ) are specified, the GS equation is solved using the Green function integral method. The poloidal magnetic flux $\psi$ at the point $(R, Z)$ produced by a unit toroidal current filament located at $(R', Z')$ is the Green function $(R', Z'; R, Z)$, which can be calculated as follows [16],

$$G(R',Z';R,Z) = \frac{\mu_0}{\pi k}\sqrt{RR'}\left\{\left(1 - \frac{k^2}{2}\right)K(k) - E(k)\right\} \qquad (4)$$

where $E(k)$ and $K(k)$ are the elliptic integrals of the first kind, and $k = \sqrt{4RR'/\{(R + R')^2 + (Z + Z')^2\}}$. The distribution of $\psi(R,Z)$ can then be obtained by convoluting G with the source term (1) over the distributed current region, as:

$$\psi(R,Z) = \int G(R',Z':R,Z) J_\phi(R',Z') dA \qquad (5)$$

The GS equation solver computes the poloidal flux function and toroidal current density on a grid point of rectangular meshes with a size of 1 cm by 1 cm, that maps the RELAX poloidal cross-section into about 2000 elements. The close-fitting conducting shell imposes a boundary condition of constant $\psi$ on its surface. In order to impose at the wall a constant flux boundary condition, 157 poloidal distributed toroidal current filaments are assigned on the inner surface of the wall, the magnitudes of which are adjusted to satisfy the boundary condition]. The computation of $\psi$ is performed by converting Eq.(5) into matrix formula, including the contribution to $\psi$ from the current filaments on the wall (see Ref.[13] for more details).

The functional form of *F* is specified by several parameters with a basis function expansion as in Ref.[13]. The functional form of *p* instead is given by three free parameters as $p = p_0 \left(1 - \psi^\beta\right)^\alpha$, where $p_0$ is the pressure on axis, with $\alpha$ and $\beta$ parameters representing broadness and peaking of the profile, respectively. Finally to minimize the equilibrium reconstruction errors a downhill simplex method is used [17].

Note that, without internal measurements the errors of the profile reconstruction could be relatively large in the plasma core. However in this paper we are interested to averaged quantities over the entire cross section (F and Θ) which are much less affected by these reconstruction errors, since both the field components at the edge and the average of the toroidal field are strongly affected by the edge profiles. In addition, we are mainly looking at the relative change of these parameters in cylindrical and toroidal geometry, which is also not much influenced by a precise reconstruction of the profiles in the core.

## 3. Axisymmetric equilibrium reconstructions with VMEC

The spectral code VMEC [12], largely adopted in the magnetic fusion community for the reconstruction of generic 3D equilibria, was modified some years ago to deal with the helical states of the RFP configuration [18,19]. In any configuration featuring a null flux surface for q (or, equivalently, an infinite rotational transform), the toroidal flux $\Psi_t$ cannot be used as a radial coordinate, as it is not monotonic. For this reason, the RFP version of VMEC uses the normalized poloidal flux s=$\Psi_p$/ $\Psi_p^{wall}$ as a radial label instead of the normalized toroidal flux, as is done conversely for the description of stellarator and tokamak equilibria.

As mentioned in the introduction, the code uses the safety factor profile as an input. In the following, such profile will be obtained either from the SHR states or from the RelaxFit output in toroidal geometry. The toroidal flux at the wall is another input quantity for VMEC, useful to compare the results in physical units. Due to the uncertainties in the pressure profile, VMEC will be run in force-free approximation. The plasma boundary is assumed to be the machine wall, circular and axi-symmetric, with minor radius a=0.25 m and major radius $R_o$ =0.51 m.

As we will see in the following, the resulting VMEC equilibria exhibit strong toroidal effects for this low aspect ratio device, and in particular a rather large outward shift of the magnetic axis with an appreciable ellipticity. Here, in particular, we will investigate the role of toroidicity on the quantities F and $\Theta$, and see how the inclusion of a proper geometric description is going to shift the points on the F-$\Theta$ diagram.

To mimic the experimental measurements of RELAX, which are taken at the top and at the bottom of the shell, we similarly compute F and $\Theta$ - because of the up-down symmetry of the equilibrium model - at the top of the plasma wall, i.e., at $\theta_g = \pi/2$, being $\theta_g$ the geometric poloidal angle with respect to the center of the circular section. The VMEC angular coordinates are the geometric (cylindrical) toroidal angle $\phi$ and a non-straight-field-line poloidal angle $\theta$. From the contravariant/covariant components of B we build the measurable toroidal and poloidal fields, $B_t$ and $B_p$, respectively. The quantities we need are

F= $B_t$ (s=1,$\theta_g=\pi/2$) $\pi$ $a^2/\Psi_t^{wall}$ ; $\Theta$= $B_p$ (s=1,$\theta_g=\pi/2$) $\pi$ $a^2/\Psi_t^{wall}$.

A comparison with the cylindrical geometry is useful, in our context, to isolate the effect of toroidicity. For this reason, we calculate also an intrinsic VMEC cylindrical limit by artificially increasing the aspect ratio, i.e., by rescaling the major radius by a factor k>>1, $R_0 \rightarrow k\ R_0$, keeping the same minor radius a. Consistently, the poloidal flux is amplified by a factor k, $\Psi_p \rightarrow k\ \Psi_p$, and the safety factor is decreased accordingly, $q \rightarrow q/k$, keeping the toroidal flux unaltered. Of course, if k is sufficiently large, the field variations in the poloidal direction are suppressed, and the quantities F and $\Theta$ should match the outcome of a purely cylindrical model.

## 4. From cylindrical to toroidal equilibria

In Ref.[9] it was shown that within the uncertainties of the reconstruction procedure the axisymmetric experimental equilibria seem to be reasonably well fitted, over the whole F-$\Theta$ range, by a parallel current following the law : $\lambda(r) = \lambda(0) (1-(r/a)^4)$ with $\lambda(0)$ changing with F (or $\Theta$). Therefore we start the comparison from this assumption. We calculate the cylindrical states at different (F,$\Theta$) by varying $\lambda(0)$, and use, as initial guess, the corresponding q profiles in VMEC and calculate the new equilibria based on the RELAX geometry and the respective cylindrical limit, as described in the previous section. The results are given in Fig.1.

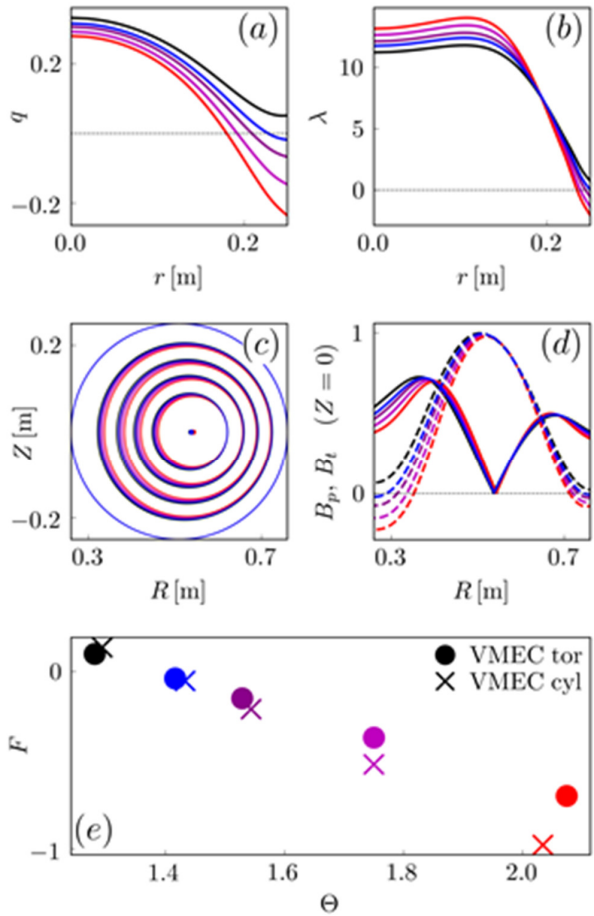

*Fig.1 : VMEC RELAX equilibria obtained with initialization from a quartic model for the parallel current (see text for details): (a) q profiles, (b) corresponding toroidal $\lambda$ profiles , (c) $\varphi$=const flux surface sections, (d) poloidal and toroidal (dashed) field magnetic field at Z=0, and (e) resulting points on the F-$\Theta$ diagram (solid dots) with the points obtained in the cylindrical limit (crosses, for R=20 $R_0$).*

It can be clearly seen that the toroidal high Θ cases are shifted to the right of the F-Θ plane (solid dots). This effect is mainly due to the poloidal and toroidal field variations along the poloidal angle, which are rather strong, as can be seen in Fig.1 (b). On the other hand, in the cylindrical limit (obtained setting k=20) the poloidal variation of the fields is effectively suppressed, and, plotted on the F-Θ plane, the resulting equilibria (represented by crosses) perfectly overlap to the pure cylindrical SHR states (as calculated from a cylindrical code).

From Fig. 1 we also note that, at high Θ, the λ profile reverses sign at the edge again because of the toroidicity.

Moving to the SHR states described by Eq.(1), as discussed in Ref.[9], the experimental data show a dependence of the single helical dominant mode on F ( or Θ). In particular at low Θ the toroidal mode number of the SH perturbation has n=3 or 4 with poloidal mode number m=1, while at higher Θ's the dominant helicity is n=6. The parallel current profile, λ(r), predicted by the relaxation model also depends on Θ and on the prescribed n value.

Also the edge current reversal was already discussed in [9], where it was artificially imposed to the λ profile, both in order to match more satisfactorily the low Θ Relax cases and especially to construct an approximate ohmic equilibrium (see section III in Ref.[9]). In addition in [9] it was shown that it was not possible to construct the n=6 relaxed states at high Θ (since the λ profile becomes extremely hollow assuming d=0,1 in Eq.(1) with a vanishing λ at the wall). Therefore, inspired by the toroidal results shown in Fig.1(d), we let the parallel relaxed current profile to reverse at the edge (by modifying the boundary conditions of Eq.(1)) also allowing the degree of reversal to increase with F (similarly to what seen in the torus. The results (with d=0,1 ) are shown in Fig.2 for a shallow reversal with n=4 (Fig.2a) and for a deeper reversal case with n=6 (Fig.2b).

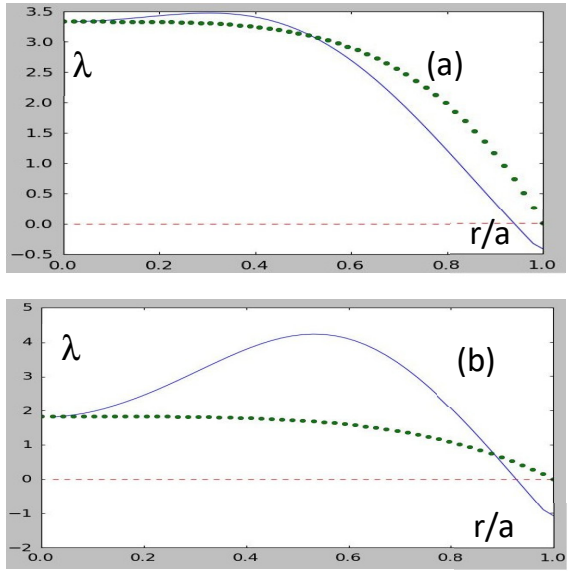

**Fig.2** $\lambda$ profiles vs. radius for: (a) n=4 dominant mode F=-0.06 $\Theta$ =1.62 , (b) n=6 dominant mode F=-1.4 $\Theta$=2.52. The dotted line is a $\lambda$ profile following a quartic: $\lambda(0) (1-(r/a)^4)$.

It can be seen that the edge current reversal allows to find an n=6 solution at deep reversal, limiting the degree of hollowness of the $\lambda$ profile, while for the n=4 case at shallow reversal the hollowness is mild and resembles more closely the trend observed in the toroidal case (see Fig.1(d)). Even for these cases with parallel current reversal we were instead unable to find solutions with n=6 at shallow F ( while the n=4 solutions exist also at deep reversal). This finding is in agreement with the experiments, where the n=6 mode is only observed at high $\Theta$. From Fig.2 it can be seen that although at low $\Theta$ the quartic profile approximate quite well the SHR $\lambda$ profile, at high $\Theta$ (Fig.2(b)) this is not true, due to the just discussed hollowness of the SHR states, which is mitigated by the assumption of the reversal of the parallel current at the edge, but it still remains quite pronounced.

An important remark is in order at this point. Although SH states in RELAX are observed at all $\Theta$'s, the persistency and robustness of them is much weaker at high $\Theta$ [20] (similarly to what seen in all RFPs at different aspect ratios). In this range of the parameters, MH states tend to prevail and SH phases emerge only for short time windows. There are several possible reasons for this to happen. One important issue is the destabilization of the m=0 modes at high $\Theta$.

Moreover experimentally there are no indication that the λ profile becomes hollow in this operational region.

Therefore, since our main purpose in this paper is to show the effect of toroidicity over cylindrical symmetry, we consider in the rest of the paper only SHR states obtained setting n=4 for the whole Θ range.

## 5. Toroidal vs. SHR cylindrical states

Next we consider in more detail a few selected cases characterized by different values of (F,Θ), reconstructed with the RelaxFit code. As in the previous analysis, the resulting q profile is given as an input to the VMEC code, then the ensuing equilibria are compared. An example can be seen in Fig.3 for two extreme F values, F=0 and F=-1 (left and right in the figure, respectively).

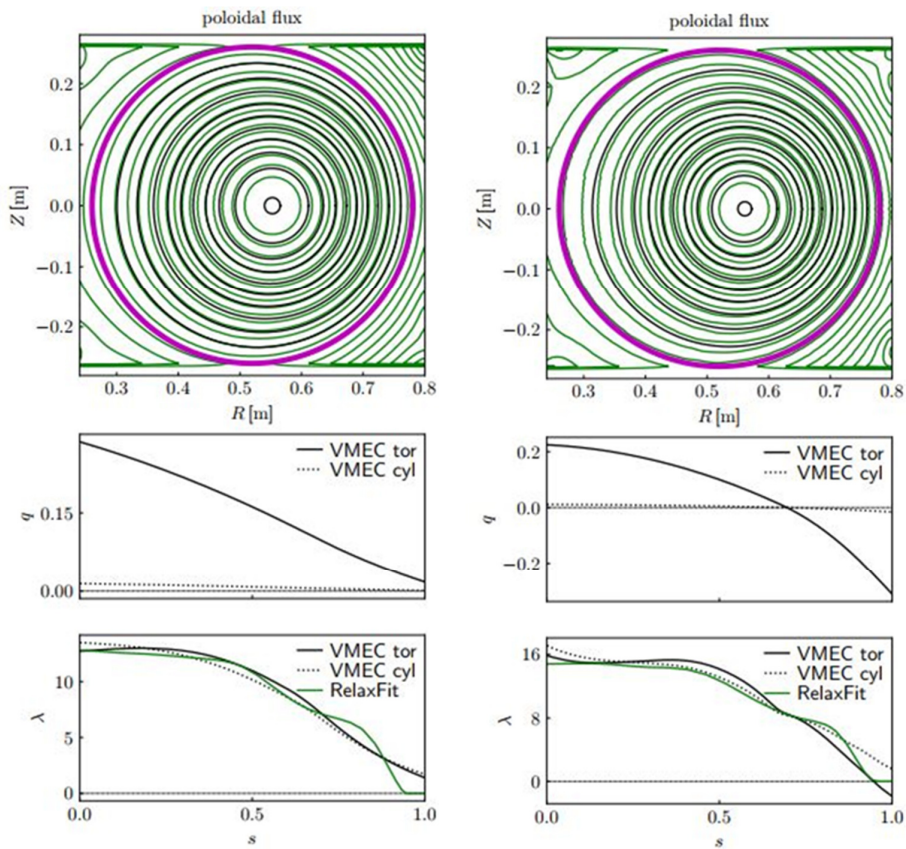

*Fig.3 Poloidal flux contours comparison (black VMEC, green RelaxFit), q and λ profiles (dashed line is the VMEC cylindrical limit with k=20), to the left a case with F=0 (case B in Fig.4) and to the right a case with F=-1 (case G in Fig.4).*

There is a quite good agreement between RelaxFit and VMEC, although the VMEC profiles appear somewhat smoother, in particular the λ profile. As in the previous section, the VMEC calculation is also done in the cylindrical limit, with k=20, showing a tendency to lose the reversal of the plasma current at the edge in the deeply reversed cases.

In Fig. 4, we show the selected (F,Θ) couples calculated with VMEC, again in the real geometry and in the cylindrical limit, together with the experimental data region. As we can see, the inclusion of a proper toroidal geometry is critical to reconcile the reconstructions with the experimental results, especially in the intermediate and high-Θ range. For comparison, we also show the SHR line (orange dashed), obtained with d=1 and n=4 (as discussed in section 4). Along this line the orange diamond, corresponding to a SHR state, is moved up by the toroidal VMEC calculation (orange circle). The discrepancy between the SHR states and experimental (F,Θ) values is eliminated when toroidicity is considered. In the figure, for comparison, also the curve (blue dashed) corresponding to the quartic λ profile is shown. It can be seen that it is more far from the experimental region than the SHR states. Moreover the distance tends to increase with Θ.

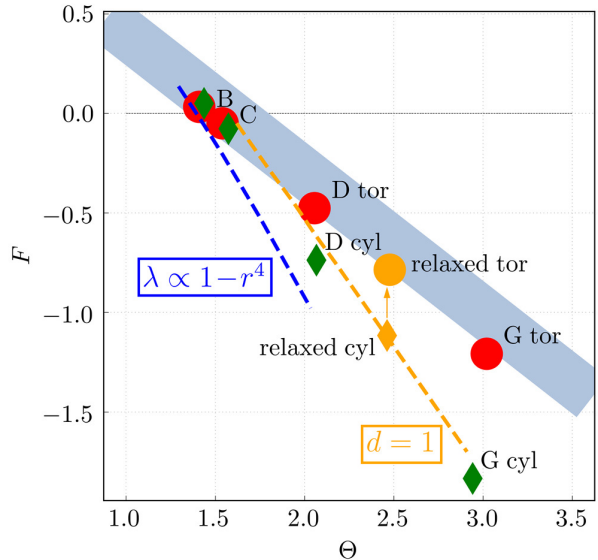

*Fig.4 F-Θ plot with the experimental region for RELAX (in light blue), toroidal VMEC equilibria (red circles) and cylindrical VMEC limit (green diamonds). The orange points represent a relaxed state in cylindrical geometry (diamond), and a VMEC equilibrium with the*

*same q profile assuming the RELAX geometry (circle). The cylindrical relaxed states are obtained with d=1 and n=4. The quartic line is also sketched in blue.*

In this case even the toroidal corrections cannot match the experimental data. For the two orange points in Fig.4 the comparison in terms of $\lambda$ profiles is also shown in Fig.5.

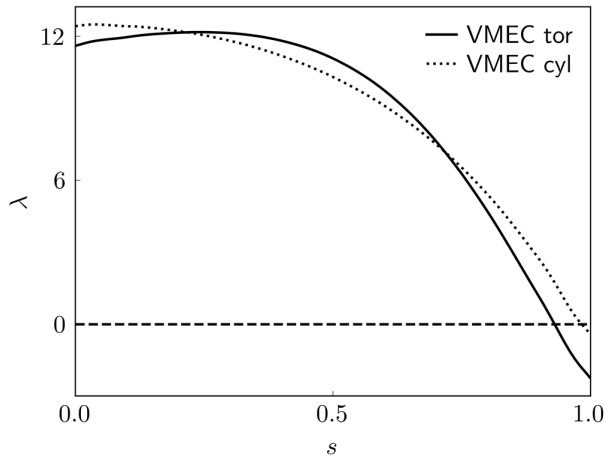

*Fig.5 $\lambda$ profiles vs the normalized poloidal flux. The dotted line corresponds to the cylindrical limit (orange diamond in Fig.4), and the plain line to the VMEC toroidal equilibrium assuming the RELAX aspect ratio (orange circle in Fig.4).*

It can be seen that the $\lambda$ profile in the toroidal case is a bit hollow and much more reversed at the edge in comparison with the cylindrical SHR case.

## Discussion and Conclusions

The topic of plasma relaxation under appropriate invariants has a long history starting from Woltjer [22] for astrophysical plasmas to Taylor [8] for the application to laboratory experiments. This fundamental and general idea has a been applied to different plasmas in nature (see for example [23-25]).

In recent years the interpretation of SH states observed in RFPs has relied on complex 3D numerical simulation which have shown the appearance, in certain regimes, of a dominant mode [26-29]. One of major limitation of the relaxation approach proposed by Taylor being, in fact, its inability to predict the correct helicity for the SH states in RFP's.

In the presence of a dominant mode of a given helicity a generalization of the relaxation theory has been proposed [10] and recently its predictions have been compared with the experimental RFP data at different aspect ratios [9,30]. As the original Taylor's approach, also this one assumes cylindrical geometry. The theory has been applied to experimental data, finding that at low aspect ratio the predictions are not good enough, at least in a certain parameters range [9]. Since the theory cannot be directly generalized to the toroidal geometry, in this paper, to test the effects of the toroidal geometry in determining the plasma operational region at low aspect ratio, we initialize the toroidal code VMEC with the q profile obtained from the SHR states.

We have shown that the poloidal asymmetries both on the poloidal and toroidal fields due to toroidicity are changing the cylindrical values of the two adimensional parameters, F and $\Theta$, towards values that agree well with the experimental measurements. The change is observed also by considering a generic cylindrical equilibrium, for example the quartic $\lambda$ profile used in section 4. However as shown in section 5 the toroidal correction could hardly match the experimental data in this case, if $\Theta$ is larger than 2.

We have also benchmarked the RelaxFit and VMEC codes showing that they give very similar results once that VMEC is initialized with the RelaxFit q profile as calculated from the magnetic equilibrium after inverse reconstruction from a set of magnetic measurements.

Since by initializing VMEC with the SHR q profiles, the F-$\Theta$ values move towards the experimental data region, we can conclude that the toroidicity is a key element to reconcile the SHR intermediate and high $\Theta$ cases with the experimentally measured F - $\Theta$ values.

An important remark is however, as discussed also in section 4, that the applicability of the relaxation model at high $\Theta$ may be questionable, since there are much less experimentally observed SH states in this parameter region, due to the destabilization of different modes, in particular the m=0, which can become dominant [21]. It is well known that a large reversal of the toroidal field can linearly destabilize these modes in RFPs, however a more subtle effect can be directly related to the development of the SH states in a torus. Since the SH has poloidal mode number m=1, it could be expected that also m=0 and m=2 toroidal sidebands are excited. The amplitude of these sidebands scales like the inverse aspect ratio and therefore would be particularly strong in RELAX. This could explain the difficulty of obtaining good SH states in this low aspect ratio device at high Θ (i.e. deep reversal). In fact a growing SH mode can more effectively, through its toroidal m=0 sideband, destabilize the already linearly unstable m=0 modes. This effect is slightly mitigated in larger aspect ratio devices, where the toroidal sidebands are smaller in amplitude. The mechanism would be very similar to the effect of an error fields on unstable MHD modes. It remains in any case true, as we have shown here, that by initializing a toroidal code like VMEC with the SHR profiles at high Θ, allows a good matching with the measurements, and therefore a better reconstruction of the plasma internal profiles, also in this, more problematic, parameter region.

A further interesting point that emerged in our study, is that the toroidal VMEC equilibria show a reversal of the parallel edge current especially at high Θ. This fact goes in the direction needed to satisfy the so called "ohmic constraint" (see [9,30] and references therein). Although in axi-symmetry this toroidal effect (as we have preliminary verified) appears still not compatible with the ohmic constraint. It remains to be checked if considering a helical toroidal equilibrium with VMEC (or with other 3D codes [28]) can help to this end. This will be also the subject of a future dedicated work.

While the local effects on the ohm's law could be appreciable, it should be noted, that the changes on the F and Θ parameters introduced by a plasma helical deformation would be very small, since δB/B (where δB is the amplitude of the helical component and B is the mean field) as measured in experiments, is only about a few 2-5 % (and is concentrated in the core being very small at the plasma edge), while the toroidal corrections, considered in this work, are much larger (as can be seen in Fig.1 from the modulation of the field components along the poloidal angle). Therefore toroidicity is certainly the strongest effect in determining the F-Θ values and their relative changes with respect to the cylindrical geometry.

Within the SHR model, it is also interesting to remark that, letting the parallel current to reverse at the edge, has allowed us to find cylindrical relaxed states with n=6 at high Θ, in correspondence to a hollow parallel current profile. This was not possible by assuming an exactly vanishing parallel current at the plasma edge as in previous calculations [9], since the λ profile was proven to become unrealistically hollow at high Θ and in some cases convergence was even not possible. Although, for the reasons discussed above, the appearance of the n=6 dominant mode at high Θ is experimentally quite seldom and not robust in RELAX, still the possibility of existence of SHR states with n=6 and edge current reversal is an interesting theoretical result.

Finally toroidal effects seem important in order to construct helical equilibria satisfying the Ohm's law. In turn, the achievement of such a state would be important in the attempt to reduce the RFP turbulence, since a helical equilibrium would not require an energy dissipation channel related to the dynamo sustainment mechanism. At low aspect ratio, this would be likely possible only at shallow reversal.